\documentclass[prl,twocolumn]{revtex4}%
\usepackage{amsmath}
\usepackage{amsfonts}
\usepackage{amssymb}
\usepackage{graphicx}%
\begin{document}
\title{Infrared signatures of hole and spin stripes in La$_{2-x}$Sr$_{x}$CuO$_{4}$}
\author{W.J. Padilla}\altaffiliation{present
address: Los Alamos National Laboratory, MS K764 MST-10, Los
Alamos NM 87545} \email{willie@lanl.gov}
\author{M. Dumm}\altaffiliation{present
address: 1. Physikalisches Institut, Universit\"{a}t Stuttgart,
70550 Stuttgart, Germany} \affiliation{Department of Physics,
University of California at San Diego, La Jolla, CA 92093-0319.}
\author{Seiki Komiya}
\author{Yoichi Ando}
\affiliation{Central Research Institute of Electric Power Industry, Komae, Tokyo 201-8511, Japan.}
\author{D. N. Basov}
\affiliation{ Department of Physics, University of California at
San Diego, La Jolla, CA 92093-0319.}

\date{\today}

\begin{abstract}
We investigate the hole and lattice dynamics in a prototypical
high temperature superconducting system
La$_{2-x}$Sr$_{x}$CuO$_{4}$ using infrared spectroscopy. By
exploring the anisotropy in the electronic response of CuO$_2$ planes we
show that our results support the notion of stripes. Nevertheless,
charge ordering effects are not apparent in the phonon spectra.
All crystals show only the expected infrared active modes for
orthorhombic phases without evidence for additional peaks that may
be indicative of static charge ordering. Strong electron-phonon
interaction manifests itself through the Fano lineshape of several
phonon modes. This analysis reveals anisotropic electron-phonon
coupling across the phase diagram, including superconducting
crystals. Due to the ubiquity of the CuO$_{2}$ plane, these
results may have implications for other high T$_c$
superconductors.
\end{abstract}

\maketitle

The nature of spin and/or charge stripes in the cuprates and their
involvement to high temperature superconductivity are currently at
the center of a debate in condensed matter physics.\cite{kivelson}
The expression ``stripes" is a general term indicating that spins
and/or holes may arrange themselves in quasi-one dimensional, or
more complicated self-organized patterns. The stripe-ordered state
minimizes the energy of holes doped in an antiferromagnetic matrix
thus leading to new inhomogeneous state of matter.\cite{goodrev}
Static one-dimensional (1D) charge stripes have been observed in
the La$_{2-x-y}$Nd$_{y}$Sr$_{x}$CuO$_{4}$ (LNSCO)
system\cite{tranquada} with complimentary evidence from neutron
and x-ray diffraction techniques. Although signatures of stripes
by way of charge ordering have not been observed in
La$_{2-x}$Sr$_{x}$CuO$_{4}$ (LSCO) and other high temperature
superconductors, there is evidence of the possible existence of
dynamical stripes in these systems.\cite{yamada} On the other
hand, spin stripes have been discovered in LSCO\cite{dynamic} and
many researchers agree upon their universality in high-T$_{c}$
superconductors. However a key issue regarding this new electronic
state of matter concerns the role of stripes in relation to
superconductivity, i.e. whether they are responsible for high
temperature superconductivity or a competing phase with it.

Signatures of quasi one dimensional behavior should be observable
in optical spectroscopy. In particular the lowering of symmetry
due the formation of rigid charge density waves results is known
to have dramatic implications for infrared (IR) active
phonons.\cite{gruner} Electronic 1D behavior can be directly
probed by way of frequency dependent conductivity. An observation
of the anisotropic conductivity in weakly doped LSCO is consistent
with the notion of stripes \cite{ando02,dumm}. Notably these
observations imply deviations from a hypothetical model in which
the stripes are rigid "rivers of charge" embedded within an
antiferromagnetic background. Moreover, electronic and lattice
fingerprints of 1D transport have not been systematically explored
as a function of doping. The goal of this work is to apply
infrared spectroscopy for the purpose of a detailed examination of
spin/charge ordering effects in a series of well characterized
LSCO crystals.

Infrared spectroscopy is a mature and powerful technique for the
study of phonons\cite{marel,padilla} electron-lattice
coupling\cite{davis,lupi} anharmonicity,\cite{cowley} and charge
and spin ordered states\cite{gruner} resulting from a lowering of
electronic and magnetic symmetry. Since precise measurements can
be carried out with miniature single crystals, IR spectroscopy
further permits a survey of the electronic anisotropy, which is
naturally expected due the formation of stripes in cuprates. The
inherent frequency dependent nature of IR spectroscopy coupled
with temperature dependent measurements, allows one to determine
characteristic energy scales associated with stripe formation as
well as their T-dependence.

\begin{figure}
[ptb]
\begin{center}
\includegraphics[
width=3.25in,keepaspectratio=true
]%
{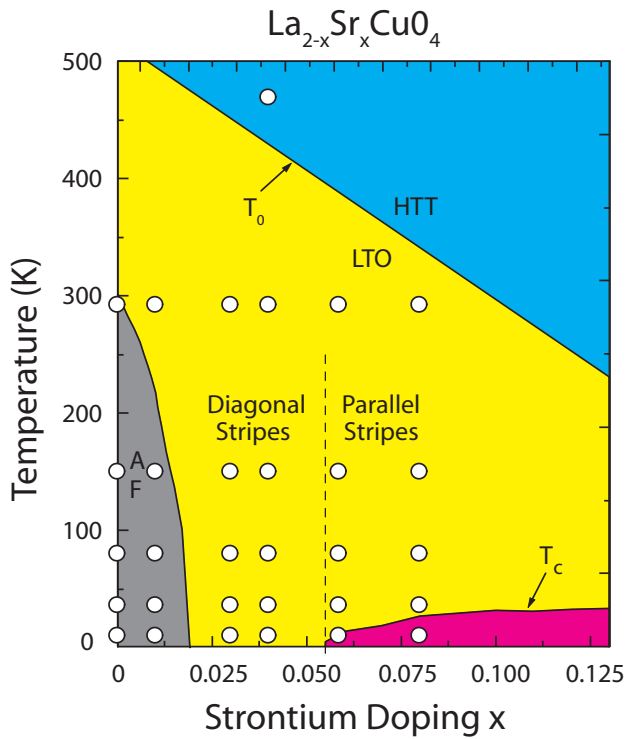}%
\caption{Phase diagram of LSCO. Circles indicate samples and
temperatures characterized in this study. AF refers to the long
range 3D antiferromagnetic regime. The superconducting dome is
indicated by the temperature T$_{c}$. The crystal structure HTT
$\rightarrow$ LTO is also noted.}%
\label{fig1}%
\end{center}
\end{figure}

\begin{figure}
[ptb]
\begin{center}
\includegraphics[
width=3.25in,keepaspectratio=true
]%
{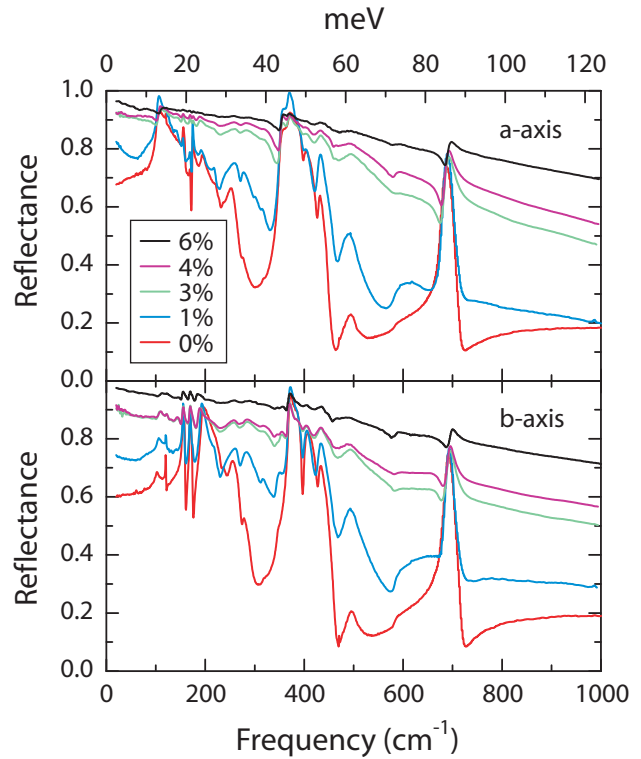}%
\caption{Infrared region of the low temperature (10 K) reflectance
for all detwinned crystals characterized. The a-axis spectra are
displayed in the top panel and the b-axis results are shown in the
bottom panel. The undoped parent compound (red curves) reveals its
insulating character, and several IR active phonons can be
observed in both axes. As carriers are added the reflectance
background increases
monotonically to larger values.}%
\label{fig2}%
\end{center}
\end{figure}

Here we present a systematic in-plane spectroscopic investigation
on a series of detwinned (0 $\leq$ x $\leq$ 0.06) single crystals
and twinned (x = 0.08) crystals of La$_{2-x}$Sr$_{x}$CuO$_{4}$,
see Fig. \ref{fig1}. From the parent Mott insulator to the
strontium doped superconducting samples (x = 0.06, 0.08) the in
plane electrodynamics are characterized by reflectance
measurements using polarized light. Samples were coated \textit{in
situ} with gold or aluminum, and spectra measured from the coated
surface were used as a reference. This method, discussed
previously in detail,\cite{KK} allows one to reliably obtain the
absolute value of the reflectance by minimizing the errors
associated with nonspecular reflection and small sample size. The
optical conductivity $\sigma_{1}(\omega)+i\sigma_{2}(\omega)$\ and
complex dielectric function
$\varepsilon_{1}(\omega)+i\varepsilon_{2}(\omega)$\ was determined
from a Kramers-Kronig (KK) transformation of the reflectance data
after extrapolation to low and high energies.\cite{wooten} The
samples are high quality single crystals grown by the
travelling-solvent floating-zone technique,\cite{Ando1} and are
cut into rectangular platelets with edges along the orthorhombic
axes, (typical size of 1.5 x 0.5 x 0.1 mm$^{3}$), where the c-axis
is perpendicular to the platelets within an accuracy of
1$^{\circ}$, as determined by x-ray Laue analysis.

\begin{figure}
[ptb]
\begin{center}
\includegraphics[
width=3.25in,keepaspectratio=true
]%
{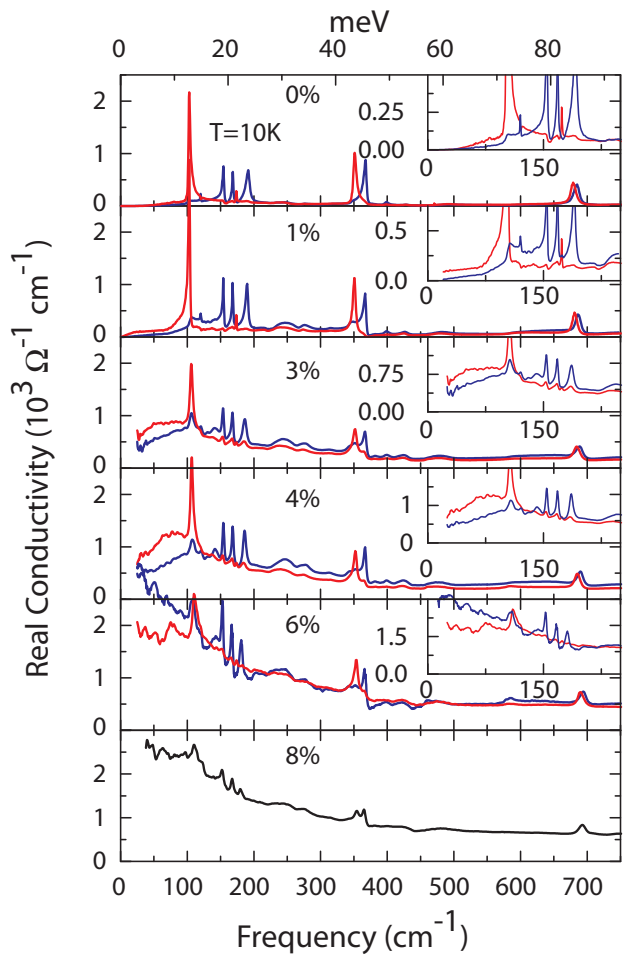}%
\caption{ The infrared portion of the real conductivity for
undoped x=0.00, top panel, and x=0.01, 0.03, 0.04, 0.06, and 0.08
in the bottom panel. In the undoped compound (top) the a-axis data
(red curves) display the expected four infrared active phonons and
the b-axis spectra (blue curves) show the expected 7 IR active
phonons. Insets detail the low frequency region and have the same
horizontal and vertical units. All data are at T=10 K.}%
\label{fig3}%
\end{center}
\end{figure}

\begin{figure}
[ptb]
\begin{center}
\includegraphics[
width=3.25in,keepaspectratio=true
]%
{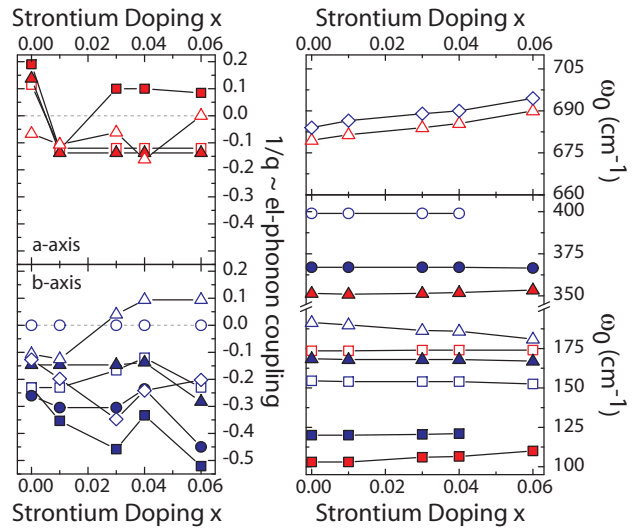}%
\caption{ Left panels: the doping dependence of the Fano parameter
(1/q) which is proportional to the electron phonon coupling, see
text for a detailed description. Right panels: dependence of the
center frequency of each IR active phonon vs. doping. The symbols
in the right panel denoting each phonon may be
used as a key for the left panel. All data are at T=10 K.}%
\label{fig4}%
\end{center}
\end{figure}

One novelty of this study compared to existing results is the
utilization of detwinned single crystals. Previous studies used
ceramic samples or twinned crystals which make the results
inconclusive or hard to interpret. The possibility to produce
detwinned single crystals of LSCO has been known for over a
decade.\cite{Thio1} However, well-detwinned LSCO crystals have
become available only after the recent finding that twin
structures in these materials can be observed on a polished
ac-face under a microscope.\cite{ando02, lavrov} Hence there has
yet to be a systematic investigation of the in-plane anisotropy as
a function of temperature. A common shortcoming of existing
studies is the lack of observation of all of the theoretically
expected phonons in the LTO (low temperature orthorhombic) phase
of LSCO. Some infrared studies of LSCO have supposed the reason
for this was due to the fact that the transition to
orthorhombicity was small\cite{tajima} and had negligible effects
on phonon/electronic structure, or that the presence of screening
currents due to carriers obstructed their
observation.\cite{Litvinchuck} As will be shown, data for
untwinned single crystals reported here allow us to remedy
problems with earlier experiments.

It is instructive to discuss the phonon features of LSCO in the
context of the evolution of the crystal structure as a function of
temperature and doping. The phase diagram plotted in  Fig.
\ref{fig1} is specific to LSCO, but many of the trends are generic
for the Cuprates. LSCO has two distinct crystallographic phases, a
high temperature tetragonal (HTT) phase, and a low temperature
orthorhombic (LTO) phase. The separation between these two phases
is denoted with the T$_0$ line in Fig. \ref{fig1}. The undoped
crystal La$_{2}$CuO$_{4}$ (La214) is a Mott insulator and exhibits
long range 3D antiferromagnetism (AF). As holes are added to the
system by substitution of Sr$^{3+}$ for La$^{2+}$, the AF quickly
dies off and a superconducting phase is found for dopings x $\geq$
0.055. The dashed vertical line in Fig. \ref{fig1} indicates the
region where stripes are know to undergo a 45$^{\circ}$
rotation.\cite{Matsuda} The open circles indicate each doping and
temperature characterized in this study.

Insights into the expected number of IR active phonon modes in the
LSCO system are provided by point group theory. LSCO belongs to a
family of structures related to the K$_{2}$NiF$_{4}$ with space
group I4/mmm (HTT) where the HTT phase can evolve into any of its
Landau subgroups Bmab (LTO), Pccn (low-T tetragonal, LTT) and
P42/ncm (low-T orthorhombic 2, LTO-2).\cite{axe} In LSCO the
HTT$\rightarrow$LTO transition is known to be of second order and
occurs as a result from the bond length mismatch between the
CuO$_{2}$ planes and the La$_{2}$O$_{2}$ bi-layers. This mismatch
is relieved by a buckling of the CuO$_{2}$ plane and a rotation of
the CuO$_{6}$ octahedra, (depicted in the lower panel Fig.
\ref{fig6}), around the a$_{ortho}$ axis which is the reason for
the structural change. Additionally since the CuO$_{6}$ octahedra
are tilted along the longer b$_{ortho}$ axis, a reduction of
symmetry compared to the HTT phase occurs and group theory
predicts 17 infrared active phonons, $\Gamma$$_{IR}$ = 6B$_{1u}$ +
4B$_{2u}$ + 7B$_{3u}$, i.e. 7 modes along the b-axis, 4 modes
along the a-axis and 6 along the c-axis. In contrast the HTT phase
has $\Gamma$$_{IR}$ = 3A$_{2u}$+4E$_{u}$, 4 modes parallel to the
ab-plane and 3 perpendicular to it. From this point on in the
manuscript we drop the ``ortho" subscript notation as all axes
referred to will be the LTO phase unless otherwise indicated.

Fig. \ref{fig2} displays the doping dependence of the reflectance
[R($\omega)$] at T=10 K for both axes in the IR regime. In the
undoped compound (red curves) relatively low values of R($\omega$)
reveal its insulating nature and oscillations, due to IR active
phonons, are observable in both axes and in all dopings studied.
As holes are added to the Cu-O plane the overall reflectance
increases smoothly and monotonically. At 6\% doping (black
curves), all phonon modes are still observable, and the high
R($\omega$) values are indicative of metallic behavior.

In Fig. \ref{fig3} we show the low temperature infrared portion of
the frequency dependent real conductivity, calculated from a KK
transformation of the data displayed in Fig. \ref{fig2}. In each
panel the strontium doping is indicated, with the undoped parent
compound La214 in the top panel and its superconducting
counterparts x=6\%, (T$_{c}$=8K) and x=8\%, (T$_{c}$=14K) in the
lower panels. The undoped parent compound exhibits low values of
the conductivity consistent with its Mott insulating behavior.
However the spectrum is punctuated by several strong phonons
visible in both the a-axis and b-axis. As expected by point group
analysis we find 4 phonons in the a-axis data at 103, 173, 351,
and 679 cm$^{-1}$ and 7 phonons in the b-axis conductivity at 120,
155, 168, 191, 367, 399, and 684 cm$^{-1}$. The novelty of this
result is that all phonons expected from group theory analysis can
be identified in the spectra for detwinned samples. Moreover, the
complete set of phonon peaks is also found in the data for twinned
superconducting crystals x=0.08 (bottom panel of Fig. \ref{fig3}).

\begin{figure}
[ptb]
\begin{center}
\includegraphics[
width=3.25in,keepaspectratio=true
]%
{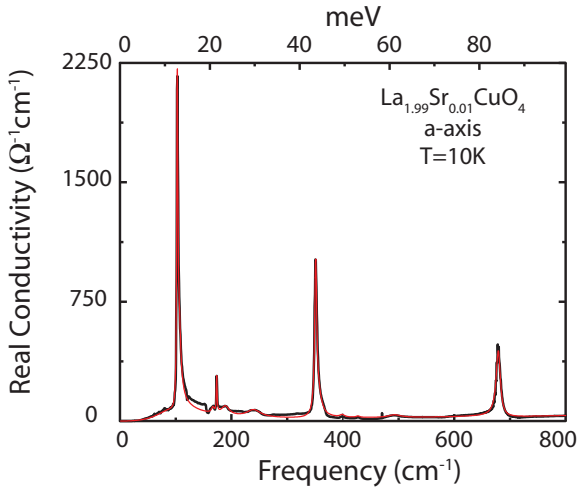}%
\caption{Frequency dependent real conductivity $\sigma_1(\omega)$
for 1\% doped LSCO in the infrared regime (black curve). The
coupling of the IR active phonons to an electronic continuum,
evident here from their asymmetry, is visible for all 4 phonons.
The red curve is a fit utilizing the Fano form for a resonance,
listed in Eq. 1.}%
\label{fig5}%
\end{center}
\end{figure}

We now turn to the analysis of the lineshape of phonon modes. Many
resonances reveal an asymmetric line shape, (for an example see
Fig. \ref{fig5}). It is instructive to quantify the degree of
asymmetry through the so-called Fano analysis. We model the
asymmetric features of the real conductivity using the form given
in Equation 1,

\begin{equation}\label{eq1}
\sigma_1 = {\frac {S\omega}{4\pi}} \left( {\frac { \left(
q\gamma+\omega-\omega_{{0}} \right) ^{2}}{{\gamma}^{2}+ \left(
\omega-\omega_{{0}} \right) ^{2}}}-1 \right)
\end{equation}
where, S is the strength of the oscillator, $\gamma$ is the
damping, and $\omega_0$ is the center frequency of the phonon. An
asymmetric phonon line shape is known to be related to
electron-phonon coupling and is typically characterized through
the Fano-Breit-Wigner (FBW) parameter q. The physical meaning of
the parameter q is that it is inversely related to the strength of
the interaction.\cite{fano} The asymmetric line shape is general
and Fano derived a form for an oscillator coupled to an electronic
continuum. These ideas were extended in ref.\cite{davis} to
include the interaction with a number of discrete states with that
of a continua. The asymmetry in the line shape determines the
energy range that the resonance is coupled. For example a line
shape which dips on the low frequency (energy) side of the
resonant frequency $\omega_{0}$, indicates the phonon is
interacting with a continuum at higher energies, where the sign of
1/q is positive in this case. For 1/q=0 a phonon assumes a
symmetric lineshape and a Lorentz form is recovered.

\begin{figure}
[ptb]
\begin{center}
\includegraphics[
width=3.25in,keepaspectratio=true
]%
{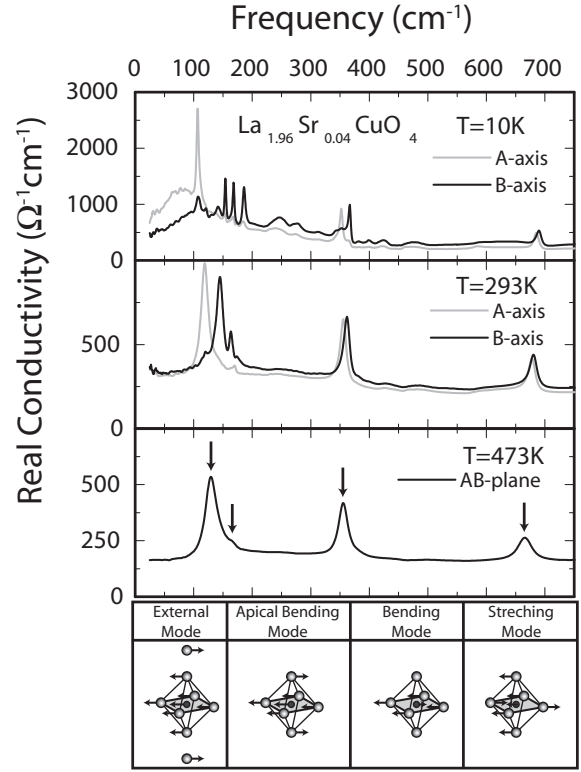}%
\caption{Infrared region of  $\sigma_1(\omega)$
La$_{1.96}$Sr$_{0.04}$CuO$_4$ crystals at 10 K, room temperature,
and in the HTT phase (T=473K). The phonons for the HTT phase are
indicated by arrows with the center frequencies given in the text.
The bottom panel lists, in order from left to right, the
assignment of each of these modes from low to high frequency as
depicted by the CuO$_{6}$ octahedra. The temperature evolution of
these modes can be tracked into the LTO phase displayed in the top
two panels. Both the a-axis (grey) and b-axis (black) data in the
LTO phase
are plotted.}%
\label{fig6}%
\end{center}
\end{figure}

In the left panels of Fig. \ref{fig4} we quantify the degree of
el-ph coupling, for both the a-axis and b-axis for and all dopings
characterized at T=10K by plotting the doping dependence of 1/q
parameter in Eq. 1. For the undoped parent compound the 1/q
parameter for the a-axis phonons  is positive for all modes with
the exception of the 685 cm$^{-1}$ phonon. This result suggests
that most of the a-axis modes interact with a continuum at higher
energies whereas the 685 cm$^{-1}$ peak  interacts with a
continuum at lower energies. In contrast all b-axis phonons in
La214 interact with a continuum at lower energies.\cite{phonon} As
carriers are introduced, at 1\% doping we find an abrupt switch to
negative 1/q values for all a-axis phonons. The b-axis phonons in
the x=0.01 crystal do not show any anomalies.  It should be noted
that the 1\% crystal is the only sample characterized in the AF
regime, in which the spin lies perpendicular to the a-axis. The
el-ph coupling relevant to the a-axis spectra is essentially
doping independent throughout the rest of the underdoped regime of
the phase diagram, with the exception of the lowest energy phonon
which changes sign between 1\% and 3\%. The b-axis phonons however
show significant signs of doping dependence and their 1/q values
are roughly 2-3 times greater than those in the a-axis. Lastly we
notice that the values for 1/q in the 6\% superconducting sample
are similar to those crystals which do not exhibit SC.

\begin{figure}
[ptb]
\begin{center}
\includegraphics[
width=3.25in,keepaspectratio=true
]%
{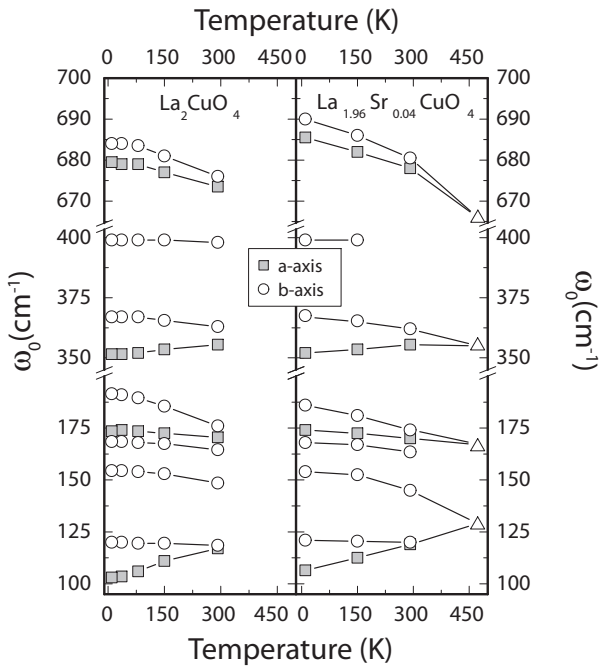}%
\caption{Temperature dependence of the optically active phonons
for undoped and 4\% doped LSCO.}%
\label{fig7}%
\end{center}
\end{figure}

Let us examine directly the HTT to LTO transition through the
spectra of the real conductivity. In Fig. \ref{fig6} the infrared
region of La$_{1.96}$Sr$_{0.04}$CuO$_{4}$ at 10K, and 292K along
with the spectrum from the HTT phase (T=473K) is shown. The HTT
phase, as stated above, has four in-plane infrared active phonons
and their normal modes are depicted in the lower portion of Fig.
\ref{fig6}, while in the panel above, each mode is indicated by an
arrow with their corresponding center frequencies at 129, 165,
355, 666 cm$^{-1}$. It can be seen that some of these modes in the
HTT phase split nearly symmetrically from their location into
phonons in the a-axis and b-axis of the LTO phase. For example the
phonon at 355 cm$^{-1}$ splits into the 352 cm$^{-1}$ and 367
cm$^{-1}$ phonons in the a-axis and b-axis respectively. The high
frequency phonon at 666 cm$^{-1}$ hardens and splits into the
phonons at 680 cm$^{-1}$ (a-axis) and 684 cm$^{-1}$ (b-axis). The
splitting is such that the a-axis phonon always lies lower in
frequency compared to the b-axis phonon in all dopings and
temperature ranges characterized, (also see the right pane of Fig.
\ref{fig7}).

\begin{figure}
[ptb]
\begin{center}
\includegraphics[
width=3.25in,keepaspectratio=true
]%
{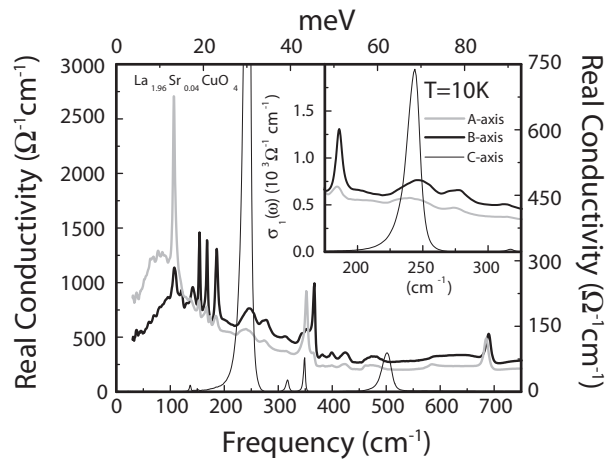}%
\caption{Low temperature infrared spectra measured for
polarization of {\bf E} vector along of all three crystallographic
axes for the LTO phase of x=4\% Sr doped LSCO. The a-axis and
b-axis spectra (thick gray and thick black lines respectively) are
shown in the main panel and their units correspond to the left
coordinate. In order to display all phonons, the c-axis
conductivity (thin black line) has been multiplied by a factor of
4 (right coordinate). The inset shows the c-axis phonon at 244
cm$^{-1}$ and the a-axis and b-axis all on the same vertical scale.}%
\label{fig8}%
\end{center}
\end{figure}

In Fig. \ref{fig7} we show the temperature dependence of the
center frequencies of all in-plane phonons for both the parent
Mott-Hubbard insulator and for 4\% LSCO. Many phonons are seen to
be temperature independent with a few exceptions. The lowest
frequency a-axis phonon at 120 cm$^{-1}$ is seen to soften with
decreasing temperature by nearly 15 cm$^{-1}$, while the b-axis
phonon at 176 cm$^{-1}$ hardens by the same amount. Phonons from
the a and b-axis of the HTT bending mode at 355 cm$^{-1}$ split
symmetrically and soften and harden, respectively, by 4 cm$^{-1}$
with the a-axis phonon lying lower in frequency. The high
frequency a-axis and b-axis phonons from the HTT stretch mode are
seen to harden by about 7 cm$^{-1}$. Interestingly they also
exhibit a significant doping dependence, as they both harden by 10
cm$^{-1}$ from undoped compound to the superconducting 6\% sample.

For completeness, we also show in Fig. \ref{fig8} the c-axis data
obtained for the x=4\% LSCO crystal along with the a-axis and
b-axis spectra, all at 10 K. As previously mentioned one would
expect 6 IR active modes in the c-axis in the LTO phase. In accord
with this expectation we find phonons at 137, 151, 244, 317, 349,
501 cm$^{-1}$. With this full characterization of lattice modes
along all crystallographic orthorhombic axes, we can understand
the origin of additional features seen for the in-plane
measurements. For example the structure near 250 cm$^{-1}$ and 500
cm$^{-1}$ in both the a-axis and b-axis spectra is likely to
originate from the mixture of the c-axis polarization. This
unwanted effect is likely to be connected with roughness of the
surface and/or minor miss-cut of the sample
surface.\cite{tajima-comment} One can also observe unwanted
"leakage" from the a-axis to the b-axis and vice versa, i.e. the
triplet of phonons centered near 170 cm$^{-1}$ in the b-axis can
be seen in the a-axis data.

Let us now examine the anisotropy of the electronic background in
the infrared conductivity data depicted in Fig. \ref{fig3}. As
holes are added to the system by increasing the strontium doping
the conductivity increases and evolves smoothly and monotonically
in both axes. We find that the electronic background shows
substantial anisotropy in all untwinned samples with dopings
$x<6\%$.\cite{ani} This is in accord with transport measurements
on detwinned single crystals of LSCO.\cite{ando02} Enhancement of
the conductivity occurs along the spin stripe direction (a-axis).
This finding is therefore consistent with the notion of inherently
conducting stripes. We also note that the anisotropy is greatest
in the 1\% crystal and is suppressed with the increasing doping.
This result supports the relevance of spin stripes for the
electronic transport in LSCO. Indeed, the separation between the
stripes is enhanced at low dopings \cite{yamada} which inevitably
should reduce the conductivity across the stripe direction.

We have shown a detailed temperature and doping dependent
systematic investigation of the electromagnetic properties of the
underdoped region of LSCO. The observation of all expected IR
active modes as predicted by group theory indicates the quality of
the samples studied here. It was found that all crystals exhibit
lineshapes suggestive of electron-phonon coupling. No anomalies in
the electron-phonon coupling are found near 5 \% doping where the
onset of superconductivity occurs in the phase diagram of LSCO. It
is therefore safe to conclude that the superconducting phase
boundary near x=5 \% is unrelated to modifications of the
interaction between the doped holes and the lattice.

Electronic anisotropy evident in Fig. \ref{fig3} supports the
notion of stripes. For example in all detwinned crystals, the low
temperature low frequency limit of $\sigma_1(\omega)$ shows that
the a-axis has larger values supporting the suggestion that
stripes are inherently conducting. The analysis of the phonon line
shape calculated in the left panels of Fig. \ref{fig4} shows
stronger electron phonon coupling along the b-axis compared to the
a-axis. This in-plane el-ph anisotropy also supports the notion of
stripes. The orthorhombicity in these systems was shown to be
small and not sufficient to account for the values calculated
here. Thus despite the two-dimensional nature of the layered
CuO$_{2}$ planes, both the electronic conductivity and phonon line
shape anisotropy are consistent with one-dimensional behavior and
therefore support the notion of stripes.

The spectra of IR active phonons detailed in Fig. \ref{fig3} show
that all expected modes evolve smoothly and continuously as
carriers are added to the system. This can be seen directly in the
right panels of Fig. \ref{fig4}. Further the temperature
dependence of the phonons vs. temperature (Fig. \ref{fig7}) shows
a continuous evolution. It must be noted that there are no
indications of new phonon modes anywhere in the underdoped region
of the phase diagram. Also the temperature and doping dependent
evolution is linear with no signs of anharmonic behavior. Since
the lowering of crystal symmetry due to the formation of charge
density waves results in a change in the number of IR modes and/or
anharmonic behavior, the findings presented above rule this out.
This result is inconsistent with the notion of static charge
ordering associated with spin stripes in LSCO. It is yet to be
seen if the hypothesis of fluctuating stripes can account for our
observations. As the CuO$_{2}$ plane is a common element to many
high T$_{c}$ superconductors, these results may be important and
more general than just limited to LSCO.

This research was supported by the U.S. Department of Energy Grant
No. DE-FG03-00ER45799.


\begin{thebibliography}{99}
\bibitem{phonon}The phonon in the b-axis located near 400cm$^{-1}$ has a small
oscillator strength which makes the determination of the FBW
constant arbitrary. Thus the asymmetric analysis has not been
carried out for this phonon, as indicated by a 1/q=0, i.e. a
Lorentzian lineshape.
\bibitem{ani}This trend is reversed in the x = 6\%
sample for which the b-axis conductivity is enhanced (bottom of
Fig. \ref{fig3}). One caveat known from neutron measurements is
that spin stripes undergo a 45$^{\circ}$ rotation at a strontium
doping of x = 5.5\%, above which the stripes lie parallel to the
Cu-O bond direction, for example see ref. \cite{Matsuda}.
Moreover, single crystals near x=5.5-6 \% doping commonly reveal
the coexistence of diagonal and collinear stripes. It is likely
that the anisotropic anomaly we find in the x=6 \% crystal may be
related to complications resulted from the coexistence of the two
forms of spin stripe patterns.
\end{thebibliography}
\end{document}